\documentclass[reprint,
               aps,
               prl,
               twocolumns,
               superscriptaddress,
               showkeys,
               showpacs]{revtex4-1}
\usepackage{graphicx}
\usepackage{hyperref}
\usepackage{booktabs}
\usepackage{subfig}
\usepackage{xcolor}

\newcommand{\et}{$E_\mathrm{T}$}

\newcommand{\kt}{$k_\mathrm{T}$}
\newcommand{\ety}{$E_\mathrm{T}^{\gamma}$}
\newcommand{\etym}{E_\mathrm{T}^{\gamma}}

\newcommand{\xf}{$x_\mathrm{F}$}
\newcommand{\xfm}{x_\mathrm{F}}

\newcommand{\mcm}{m_\mathrm{c}}
\newcommand{\gev}{GeV}
\newcommand{\tev}{TeV}

\newcommand{\fwd}{$1.56 \leq |\eta^{\gamma}| < 2.37$}
\newcommand{\cent}{$|\eta^{\gamma}| < 1.37$}
\newcommand{\chis}{$\chi^{2}$}
\newcommand{\sherpa}{\textsc{Sherpa}}

\newcommand{\wc}{$w_\mathrm{c}$}
\newcommand{\wcm}{w_\mathrm{c}}
\newcommand{\wcc}{$w_{\mathrm{c} \bar \mathrm{c}}$}
\newcommand{\wccm}{w_{\mathrm{c} \bar{} \mathrm{c}}}
\newcommand{\wul}{$w_\mathrm{u.\,l.}$}
\newcommand{\wulm}{w_\mathrm{u.\,l.}}

\newlength{\pw}
\begin{document}

\title{Constraints on the Intrinsic Charm Content of the Proton from Recent
       ATLAS Data}

\author{V.A.~Bednyakov}
\affiliation{Joint Institute for Nuclear Research, Dubna 141980, Moscow region,
             Russia}
\author{S.J.~Brodsky}
\affiliation{SLAC National Accelerator Laboratory, Stanford University,
             Stanford, CA 94025, United States}
\author{A.V.~Lipatov}
\affiliation{Joint Institute for Nuclear Research, Dubna 141980, Moscow region,
             Russia}
\affiliation{Skobeltsyn Institute of Nuclear Physics, Moscow State University,
             119991 Moscow, Russia}
\author{G.I.~Lykasov}
\email{gennady.lykasov@cern.ch}
\affiliation{Joint Institute for Nuclear Research, Dubna 141980, Moscow region,
             Russia}
\author{M.A.~Malyshev}
\affiliation{Skobeltsyn Institute of Nuclear Physics, Moscow State University,
             119991 Moscow, Russia}
\author{J.~Smiesko}
\affiliation{Comenius University in Bratislava, Faculty of Mathematics, Physics
             and Informatics, Mlynska Dolina, 842 48 Bratislava, Slovakia}
\author{S.~Tokar}
\affiliation{Comenius University in Bratislava, Faculty of Mathematics, Physics
             and Informatics, Mlynska Dolina, 842 48 Bratislava, Slovakia}

\begin{abstract}
Constraints on the intrinsic charm probability
$\wccm = P_{{\mathrm{c}\bar \mathrm{c}} / \mathrm{p}}$ in the proton are
obtained for the first time from LHC measurements. The ATLAS Collaboration data
for the production of prompt photons, accompanied by a charm-quark jet in pp
collisions at $\sqrt s = 8 $ TeV, are used. The upper limit
\mbox{$\wccm < 1.93$~\%} is obtained at the 68~\% confidence level. This
constraint is primarily determined from the theoretical scale and systematical
experimental uncertainties. Suggestions for reducing these uncertainties are
discussed. The implications of intrinsic heavy quarks in the proton for future
studies at the LHC are also discussed.
\end{abstract}


\maketitle

One of the fundamental predictions of quantum chromodynamics is the existence of
Fock states containing heavy quarks at large light-front (LF) momentum fraction
$x$ in the LF wavefunctions of hadrons~\cite{Brodsky:1980pb,Brodsky:1984nx}. A
key example is the $|uudc\bar{c}\rangle$ intrinsic charm Fock state of the
proton's QCD eigenstate generated by $c\bar{c}$-pairs which are multiply
connected to the valence quarks. The resulting intrinsic charm (IC) distribution
$c(x, Q^2)$ is maximal at minimal off-shellness; i.e., when all of the quarks in
the $|uudc\bar{c}\rangle$ LF Fock state have equal rapidity. Equal rapidity
implies that the constituents in the five-quark light-front Fock state have
momentum fractions $x_i = {k^+_i \over P^+} \propto \sqrt{m^2_i +
k^2_{\textrm{T} i}}$, so that the heavy quarks carry the largest momenta.

The study of the intrinsic heavy quark structure of hadrons provides insight
into fundamental aspects of QCD, especially its nonperturbative aspects. The
operator product expansion (OPE) predicts that the probability for intrinsic
heavy $Q$-quarks in a light hadron scales as $\kappa^2/M^2_\textrm{Q}$ due to
the twist-6 $G^3_{\mu \nu}$ non-Abelian couplings of QCD~\cite{Franz:2000ee,
Brodsky:1984nx}. Here $\kappa$ is the characteristic mass scale of QCD.\@ In the
case of the BHPS model~\cite{Brodsky:1980pb} within the MIT bag
approach~\cite{Donoghue:1977qp}, the probability to find a five-quark component
$|uudc\bar{c}\rangle$ bound in the nucleon eigenstate is estimated to be in the
range 1--2\%. Although there are many phenomenological signals for heavy quarks
at high $x$, the precise value for the intrinsic charm probability
$\wccm = P_{{\mathrm{c}\bar \mathrm{c}} / \mathrm{p}}$ in the proton has not as
yet been definitively determined. 

The first evidence for intrinsic charm (IC) in the proton originated from the
EMC measurements of the charm structure function $c(x, Q^2)$ in deep inelastic
muon-proton scattering~\cite{Harris:1995jx}. The charm distribution measured by
the EMC experiment at $x_\textrm{bj} = 0.42$ and $Q^2 = 75$~\gev{} was found to
be approximately 30 times that expected from the conventional gluon splitting
mechanism $g \to c \bar c$. However, as discussed in ref.~\cite{Hou:2017khm},
this signal for (IC) is not conclusive because of the large statistical and
systematic uncertainties of the measurement. 

A series of experiments at the Intersection Storage Ring (ISR) at CERN, as well
as the fixed-target SELEX experiment at Fermilab, measured the production of
heavy baryons $\Lambda_\mathrm{c}$ and $\Lambda_\mathrm{b}$ at high \xf{} in
$pp$, $\pi^{-} p$, $\Sigma^{-} p$ collisions~\cite{Bari:1991ty,Barlag:1990hg,
Garcia:2001xj}. For example, the $\Lambda_c$ will be produced at high \xf{} from
the excitation of the $|uudc\bar{c}\rangle$ Fock state of the proton and the
comoving $u$, $d$, and $c$ quarks coalesce. The \xf{} of the produced forward
heavy hadron in the $pp \to \Lambda_c$ reaction is equal to the sum $\xfm =
x_\textrm{u} + x_\textrm{d} + x_\textrm{c}$ of the light-front momentum
fractions of the three quarks in the five-quark Fock state. However, the
normalization of the production cross section has sizable uncertainties, and
thus one cannot obtain precise quantitative information on the intrinsic charm
contribution to the proton charm parton distribution function (PDF) from the ISR
and SELEX measurements. Another important way to identify IC utilizes the single
and double $J/\Psi$ hadroproduction at high \xf{} as measured by the NA3 fixed
target experiment~\cite{Badier:1983dg,Vogt:1991qd,Brodsky:2017ntu}.  

The first indication for intrinsic charm at a high energy collider was observed
in the \mbox{$\bar p p \to c \gamma X$} reaction at the
Tevatron~\cite{Abazov:2009de,Abazov:2012ea,D0:2012gw, Aaltonen:2009wc,
Aaltonen:2013ama}. An explanation for the large rate observed for events at high
\et{} based on the intrinsic charm contribution is given in
refs.~\cite{Rostami:2015iva,Vafaee:2016jxl}. A comprehensive review of the
experimental results and global analysis of PDFs with intrinsic charm was
presented in~\cite{Brodsky:2015fna,Ball:2014uwa}.

Intrinsic heavy quarks also leads to the production of the Higgs boson at high
\xf{} the LHC energies~\cite{Brodsky:2007yz}. It also implies the production of
high energy neutrinos from the interactions high energy cosmic rays in the
atmosphere, a reaction which can be measured by the IceCube
detector~\cite{Laha:2016dri}.

In our previous publications~\cite{Bednyakov:2013zta,Beauchemin:2014rya,
Lipatov:2016feu,Brodsky:2016fyh,Lipatov:2018oxm}, we showed that the IC signal
can be visible in the production of prompt photons and vector bosons $Z$/$W$ in
$pp$ collisions, accompanied by heavy-flavor $c$/$b$-jets at large transverse
momenta and the forward rapidity region ($|y| > 1.5$), kinematics within the
acceptance of the ATLAS and CMS experiments at the LHC.\@

The main goal of this paper is to test the intrinsic charm hypothesis
utilizing recent ATLAS data on prompt photon production accompanied by a $c$-jet
in $pp$ collisions at $\sqrt{s} = 8$~\tev. We perform this analysis using the
analytical QCD calculation and the MC generator \sherpa~\cite{Gleisberg:2008ta}. 

We will first present the scheme of our QCD analysis for such processes. The
systematic uncertainties due to hadronic structure are evaluated using a
combined QCD approach, based on the \kt-factorization
formalism~\cite{Gribov:1984tu,Catani:1990eg} in the small-$x$ domain and the
assumption of conventional (collinear) QCD factorization at large $x$. Within
this approach, we have employed the \kt-factorization formalism to calculate the
leading contributions from the ${\cal O}(\alpha \alpha_s^2)$ off-shell
gluon-gluon fusion $g^{*} g^{*} \to \gamma c \bar c$. In this way one takes
into account the conventional perturbative charm contribution to associated
$\gamma c$ production. In addition there are backgrounds from jet
fragmentation.

The IC contribution is computed using the ${\cal O}(\alpha \alpha_s)$ QCD
Compton scattering $c g^* \to \gamma c$ amplitude, where the gluons are kept
off-shell and incoming quarks are treated as on-shell partons. This is justified
by the fact that the IC contribution begins to be visible at the domain of large
$x \geq 0.1$, where its transverse momentum can be safely neglected. The
\kt-factorization approach has technical advantages, since one can include
higher-order radiative corrections by adopting a form for the transverse
momentum dependent (TMD) parton distribution of the proton (see
reviews~\cite{Andersson:2002cf} for more information). In addition, we take into
account several standard pQCD subprocesses involving quarks in the initial
state. These are the flavor excitation $c q \to \gamma c q$, quark-antiquark
annihilation $q \bar q \to \gamma c \bar c$ and quark-gluon scattering
subprocess $q g\to \gamma q c \bar c$.  These processes become important at
large transverse momenta $p_T$ (or, respectively, at large parton longitudinal
momentum fraction $x$, which is the kinematics needed to produce high $p_T$
events); it is the domain where the quarks are less suppressed or can even
dominate over the gluon density. We rely on the conventional (DGLAP)
factorization scheme, which should be reliable in the large-$x$ region. Thus,
we apply a combination of two techniques (referred as a ``combined QCD
approach{}'') employing each of them in the kinematic regime where it is most
suitable. More details can be found in~\cite{Baranov:2017tig} (see also
references therein).

According to the BHPS model~\cite{Brodsky:1980pb,Brodsky:1981se}, the total
charm distribution in a proton is the sum of the \emph{extrinsic} and
\emph{intrinsic} charm contributions.
\begin{eqnarray}
  xc(x, \mu_0^2) = xc_\mathrm{ext}(x, \mu_0^2) + xc_\mathrm{int}(x, \mu_0^2).
  \label{def:cdens_start}
\end{eqnarray}
The \emph{extrinsic} quarks and gluons are generated by perturbative QCD on a
short-time scale associated within the large-transverse-momentum processes.
Their distribution functions satisfy the standard QCD evolution equations. In
contrast, the \emph{intrinsic} quarks and gluons are associated with a
bound-state hadron dynamics and thus have a non-perturbative origin. In
Eq.~\ref{def:cdens_start} the IC weight is included in $xc_\mathrm{int}(x,
\mu_0^2)$ and the total distribution $xc(x, \mu_0^2)$ satisfies the QCD sum
rule, which determines its normalization~\cite{Blumlein:2015qcn} and see Eq.~6
in~\cite{Brodsky:2016fyh}.  We define the IC probability as the
$n = 0$ moment of the charm PDF at the scale $\mu_0 = \mcm$, where
$\mcm = 1.29$~\gev{} is the $c$-quark mass.

As shown in~\cite{Lipatov:2016feu,Brodsky:2016fyh}, the interference between the
two contributions to Eq.~\ref{def:cdens_start} can be neglected, since the IC
term $xc_\mathrm{int}(x, \mu^2)$ is much smaller than the extrinsic contribution
generated at $x < 0.1$ by DGLAP evolution~\cite{Gribov:1972ri,Altarelli:1977zs,
Dokshitzer:1977sg} from gluon splitting. Therefore, since the IC probability
$w_{c{\bar c}}$ enters into Eq.~\ref{def:cdens_start} as a constant in front of
the function dependent on $x$ and $\mu^2$, one can adopt a simple linear
relation for any $\wccm \leq \wccm^{\max}$~\cite{Lipatov:2016feu,
Brodsky:2016fyh}, which provides an interpolation between two charm densities at
the scale $\mu^2$, obtained at $\wccm = \wccm^{\max}$ and $\wccm = 0$. We have
performed a three-point interpolation of the all parton (quark and gluon)
distributions for $\wccm = 0$, 1 and 3.5~\%, which correspond to the CTEQ66M,
CTEQ66c0 and CTEQ66c1 sets, respectively~\cite{Nadolsky:2008zw}. For the
interpolation function we used the linear and the quadratic \wcc{}
interpolation. The difference between linear and quadratic interpolation
functions in the interval $0 < \wccm \leq 3.5$~\% is no greater than 0.5~\%,
thus giving confidence in our starting point~\cite{Lipatov:2016feu}. Given that,
we used the quadratic interpolation for the all parton flavors at $\mu_0$ and
$\wccm < \wccm^{\max}$ to satisfy the quark and gluon sum rules,
see~\cite{Chang:2011du,Chang:2014jba}. Let us stress that at $\wccm =
\wccm^{\max}$ the quark sum rule is satisfied automatically in the used PDF
because the intrinsic light $q{\bar q}$ contributions are
included~\cite{Nadolsky:2008zw}. Note that \wcc{} is treated in
$xc_\mathrm{int}(x, \mu_0^2)$ of Eq.~\ref{def:cdens_start} as a parameter which
does not depend on $\mu^2$. Therefore, its value can be determined from the fit
to the data.

\begin{figure}[ht!]
  \centering
  \subfloat[$\wcm = 0$~\%\label{fig1a}]{%
    \includegraphics[width=\linewidth]{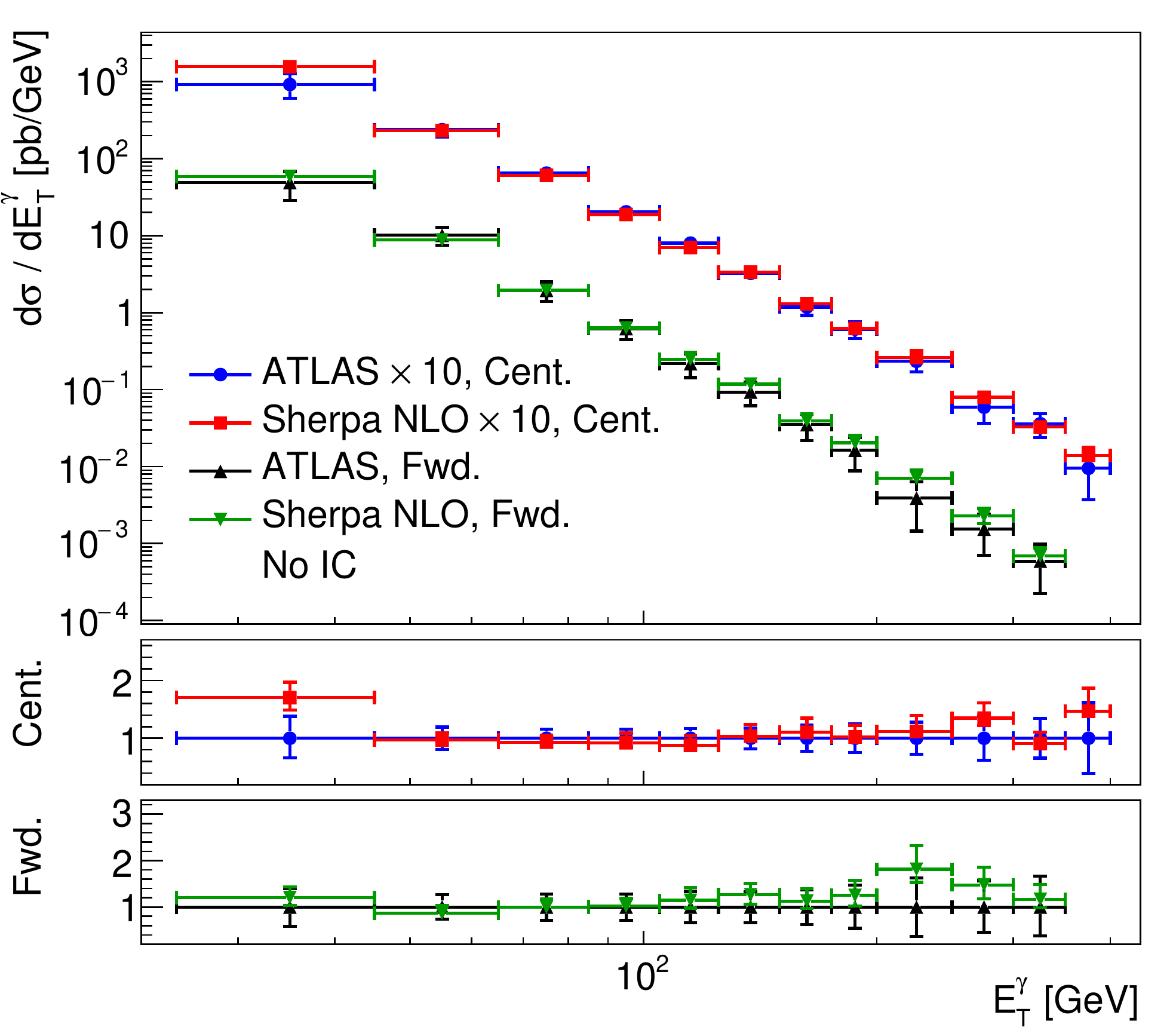}}
 
  \subfloat[$\wulm = 1.93$~\%\label{fig1b}]{%
    \includegraphics[width=\linewidth]{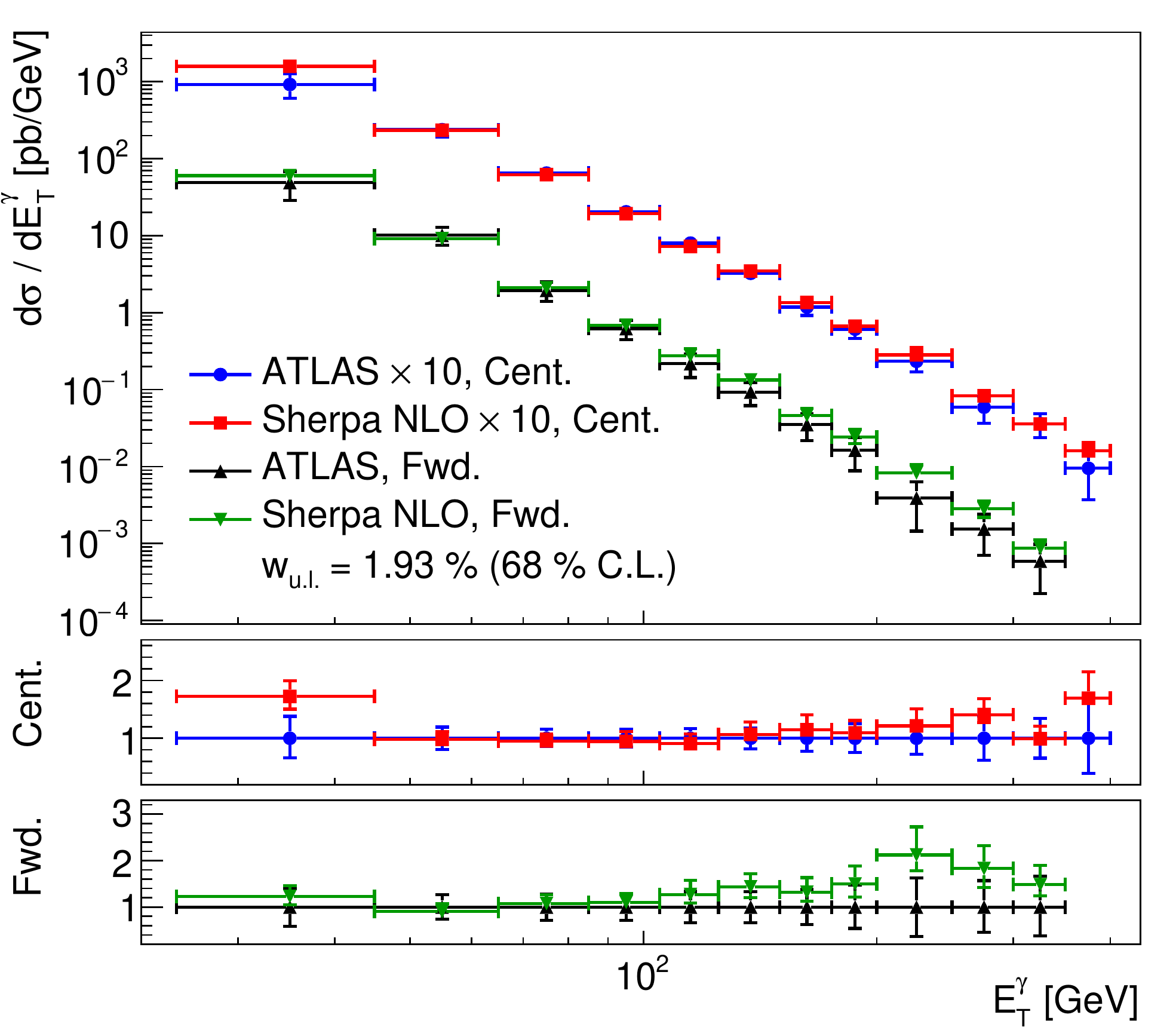}}

  \caption{The \ety-spectrum calculated with MC generator \sherpa{}, NLO 
           compared with the ATLAS
           data~\cite{Aaboud:2017skj}. \\
           (a) top: the spectrum at the central rapidity region \cent{} and
           forward \fwd{} region without the {IC} contribution; \\
           (a) middle: the ratio of the MC calculation to the data for the
           central rapidity region ($\wcm = 0$~\%); \\
           (a) bottom: the ratio of the MC calculation to the data for the for
           forward rapidity regions ($\wcm = 0$~\%). \\
           (b): the same spectra, as in (a), but with the upper limit of IC
           contribution $\wulm = 1.93$~\%.}\label{fig1}
\end{figure}

\begin{figure}[ht!]
  \centering
  \subfloat[$\wcm = 0$~\%\label{fig2a}]{%
    \includegraphics[width=\linewidth]{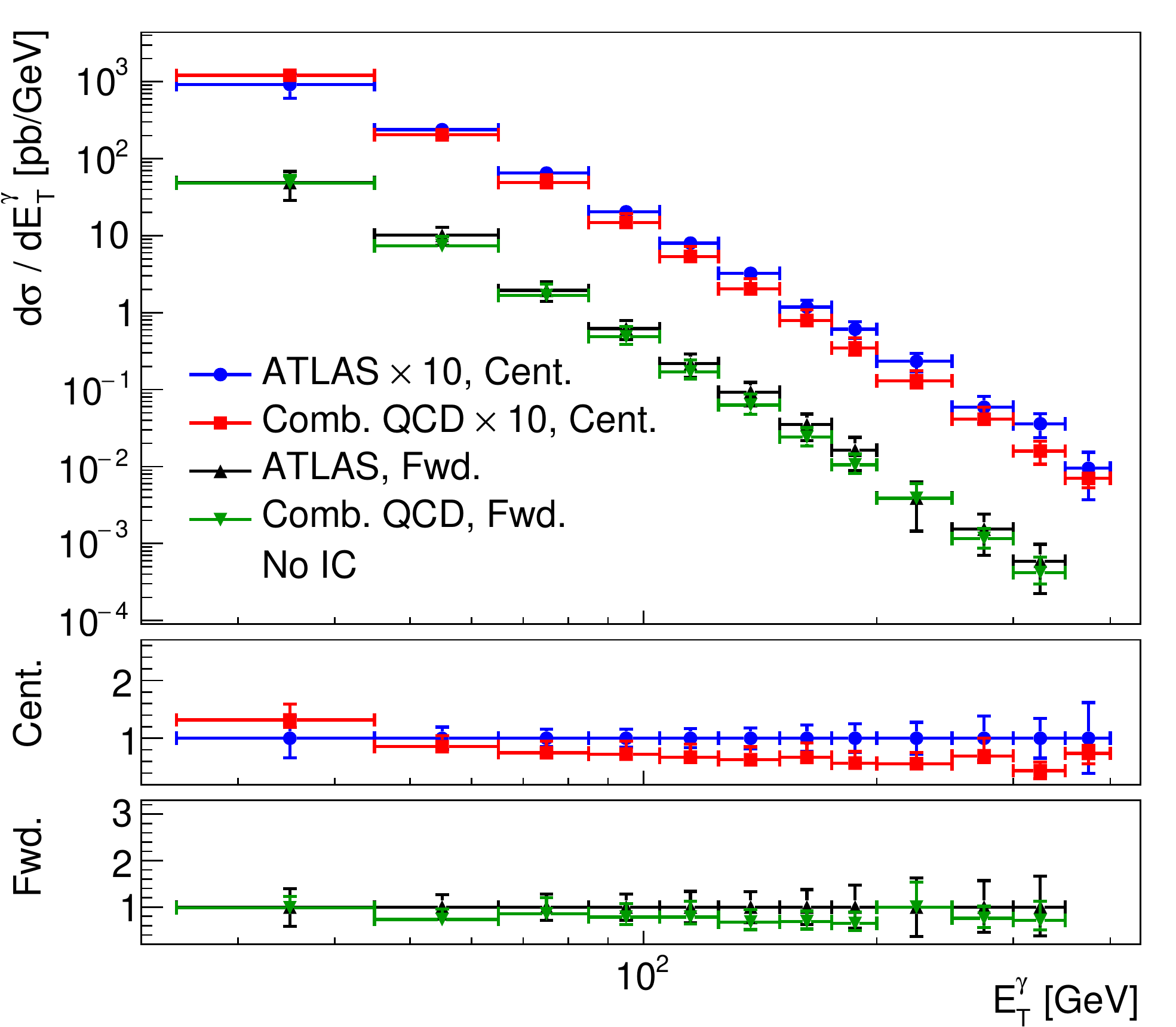}}

  \subfloat[$\wulm = 2.91$~\%\label{fig2b}]{%
    \includegraphics[width=\linewidth]{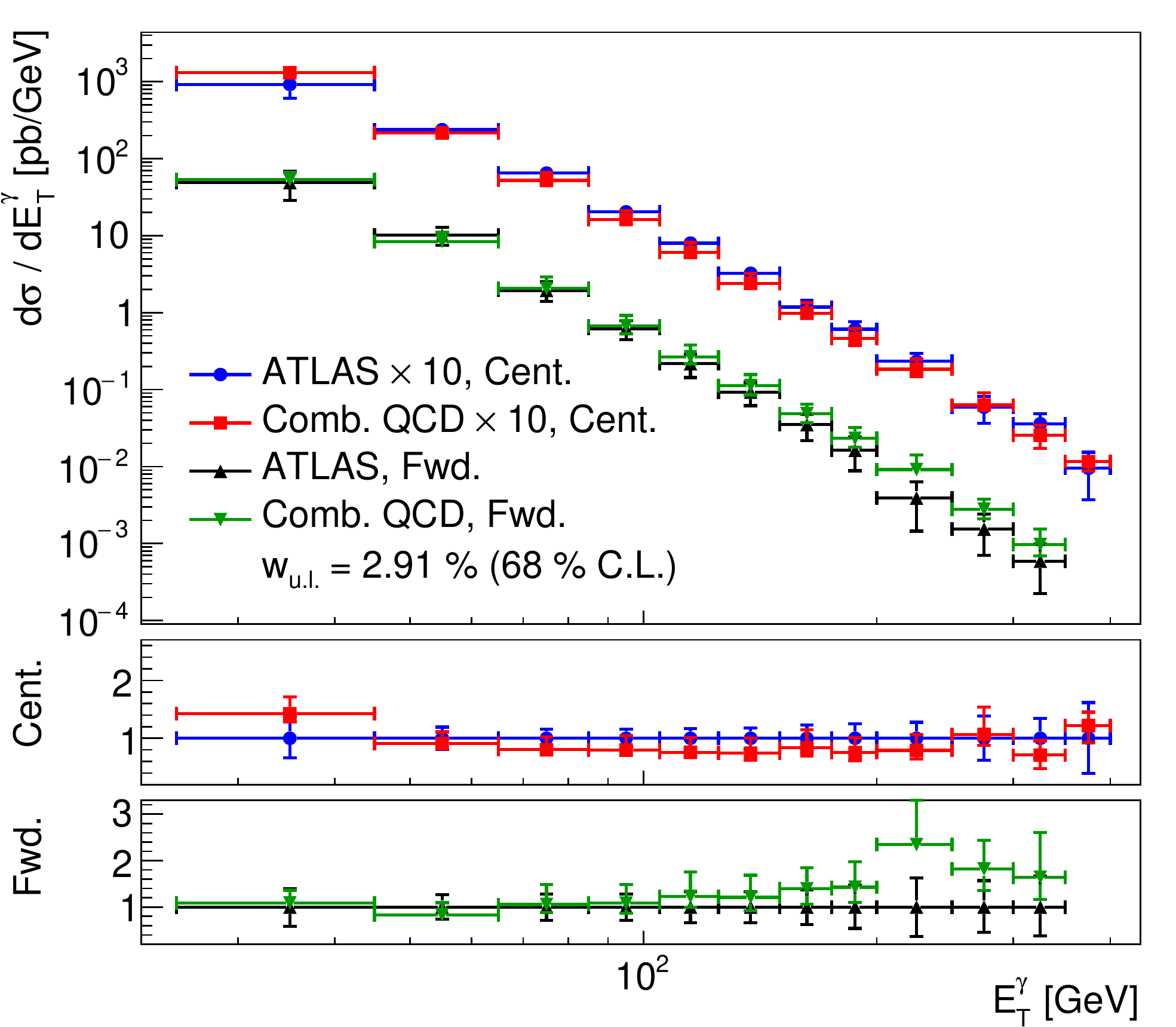}}

  \caption{The spectrum of prompt photons as a function of its transverse energy
           \ety{} calculated with the combined QCD analysis, compared with ATLAS
           data~\cite{Aaboud:2017skj}. \\
           (a) top: the spectrum in the central rapidity region \cent{} and
           forward \fwd{} region without the {IC} contribution; \\
           (a) middle: the ratio of the MC calculation to the data for the
           central rapidity region ($w_c = 0$\%); \\
           (a) bottom: the ratio of the MC calculation to the data for the
           forward rapidity regions ($w_c = 0$\%). \\
           (b): the same spectra, as in (a), but with upper limit of {IC}
           contribution $\wulm = 2.91$~\% corresponding to the best fit of the
           data.}\label{fig2}
\end{figure}

\begin{figure}[ht!]
  \centering
  \includegraphics[width=\linewidth]{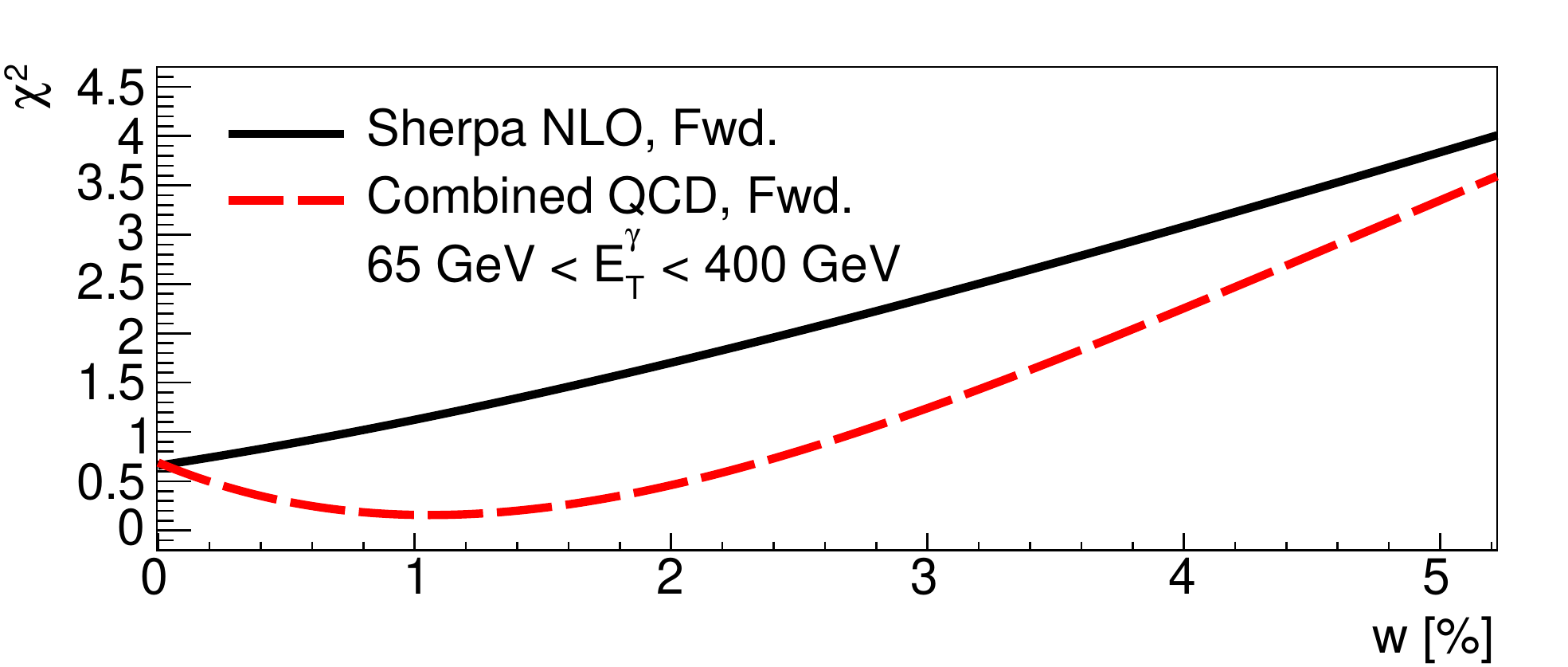}
  \caption{Solid line: \chis{} as a function of $w$ in the forward rapidity
           region in \sherpa{} NLO\@. Dashed line: Same but \chis{} obtained
           within the combined QCD calculation.}\label{fig3}
\end{figure}

For the second approach, we use the MC generator
\sherpa~\cite{Gleisberg:2008ta,Cascioli:2011va,Bothmann:2016nao} with
next-to-leading order (NLO) matrix elements (version 2.2.4) to generate samples
for the extraction of the \wcc{} from the ATLAS data. The recent versions of
\sherpa{} generator can provide additional weights, which are used to reweight
our spectra to PDFs with different IC contribution~\cite{Bothmann:2016nao}. In
order to obtain weights corresponding to any $w$ we quadratically interpolate
three PDF weights: $W_0$ (0~\% IC), $W_1$ corresponding to the BHPS1 (the mean
value of the $c\bar c$ fraction is $\langle x_{{\mathrm{c}}\bar{\mathrm{c}}}
\rangle \simeq 0.6$~\%, which corresponds to the IC probability $\wccm = $
1.14~\%) and $W_2$ corresponding to the BHPS2 ($\langle
x_{{\mathrm{c}}\bar{\mathrm{c}}} \rangle \simeq 2.1$~\%, $\wccm = 3.54$~\%). We
used CT14nnloIC~\cite{Dulat:2015mca} PDF included with the help of
LHAPDF6~\cite{Buckley:2014ana}. The process $p + p \rightarrow \gamma +$~any jet
(up to 3 additional jets) is simulated with the requirement $\etym > 20$~\gev{}
and $\eta^{\gamma} < 2.7$. Additional cuts are applied to match the ATLAS event
selection~\cite{Aaboud:2017skj}. In order to extract the $w$-value from the
data we first calculate the \ety-spectrum using the \sherpa{} MC generator in
the central rapidity region (\cent) and compare it with the $\gamma + c$-jet
ATLAS spectra. Next, we show that one can obtain a satisfactory description of
the ATLAS data in the central rapidity region using  the \sherpa{} (NLO)
calculation without IC.\@

These results using the PDF CT14nnloIC without the IC contribution are presented
in Fig.~\ref{fig1} (top). One can see that the difference between these
experimental \ety{} data and the MC calculation is less than the total
uncertainties. Therefore, we are unable to determine a precise value of the IC
probability from recent ATLAS data. However, one can extract an upper limit of
the IC contribution from the data. Therefore, the \sherpa{} NLO calculation at
the upper limit of the IC contribution $\wulm = 1.93$~\% is presented in
Fig.~\ref{fig1} (bottom). This value of \wul{} corresponds to the \chis{} at
minimum plus one, see the solid line in Fig.~\ref{fig3}. 

The \wcc{} extraction method was repeated using the above mentioned combined QCD
scheme instead of \sherpa{} (NLO). The CTEQ66c PDF,  which includes the {IC}
fraction in the proton, was used to calculate the \ety-spectrum in the forward
rapidity region \ety{} spectra and \chis{} as a function of $w$  obtained within
these approach are presented in Fig.~\ref{fig2} and Fig.~\ref{fig3} (dashed red
line) respectively. The upper limit of the IC contribution obtained within the
cobmbined QCD is about $\wulm = 2.91$~\%. As was shown
in~\cite{Lipatov:2018oxm}, the combined QCD does not include  parton showers and
hadronization, a contribution which is sizable at \ety$>$100 GeV, where the IC
signal could be visible. Therefore, the results obtained \sherpa{} (NLO), which
include these effects, are more realistic.  

The $w$-dependence of $\chi^2$-functions obtained with both \sherpa{} and the
combined QCD approach in the forward region are presented in Fig.~\ref{fig3}. By
definition, the minimum of the $\chi^2$-function is reached at a central value
\wc{} which corresponds to the best description of the ATLAS data. The
application of \sherpa{} results in $\wcm = 0.00$~\%, and the combined QCD gives
us $\wcm = 1.00$~\%.

Figure~\ref{fig3} shows a rather weak \chis-sensitivity to the $w$-value. It is
due to the large experimental and theoretical QCD scale uncertainties
especially at $\etym > 100$~\gev{} (Figs.~\ref{fig1} and~\ref{fig2}). Therefore,
it is not possible to extract the \wc-value with a requested accuracy
(3--5~$\sigma$), instead, we present relevant upper limit at the 68~\%
confidence level (C.L.).

As a first estimate of the scale uncertainty we have used the conventional
procedure, used in a literature, varying the values of the QCD renormalization
scale $\mu_\mathrm{R}$ and the factorization one $\mu_\mathrm{F}$ in the
interval from $0.5\etym$ to $2\etym$. In fact, there are several methods to
check the sensitivity of observables to the scale uncertainty, see, for
example,~\cite{Wu:2013ei} and references there in. The renormalization scale
uncertainty of the \ety{}-spectra poses a serious theoretical problem for
obtaining more precise estimate of the IC probability from the LHC data.

\begin{figure}[ht!]
  \centering
  \includegraphics[width=\linewidth]{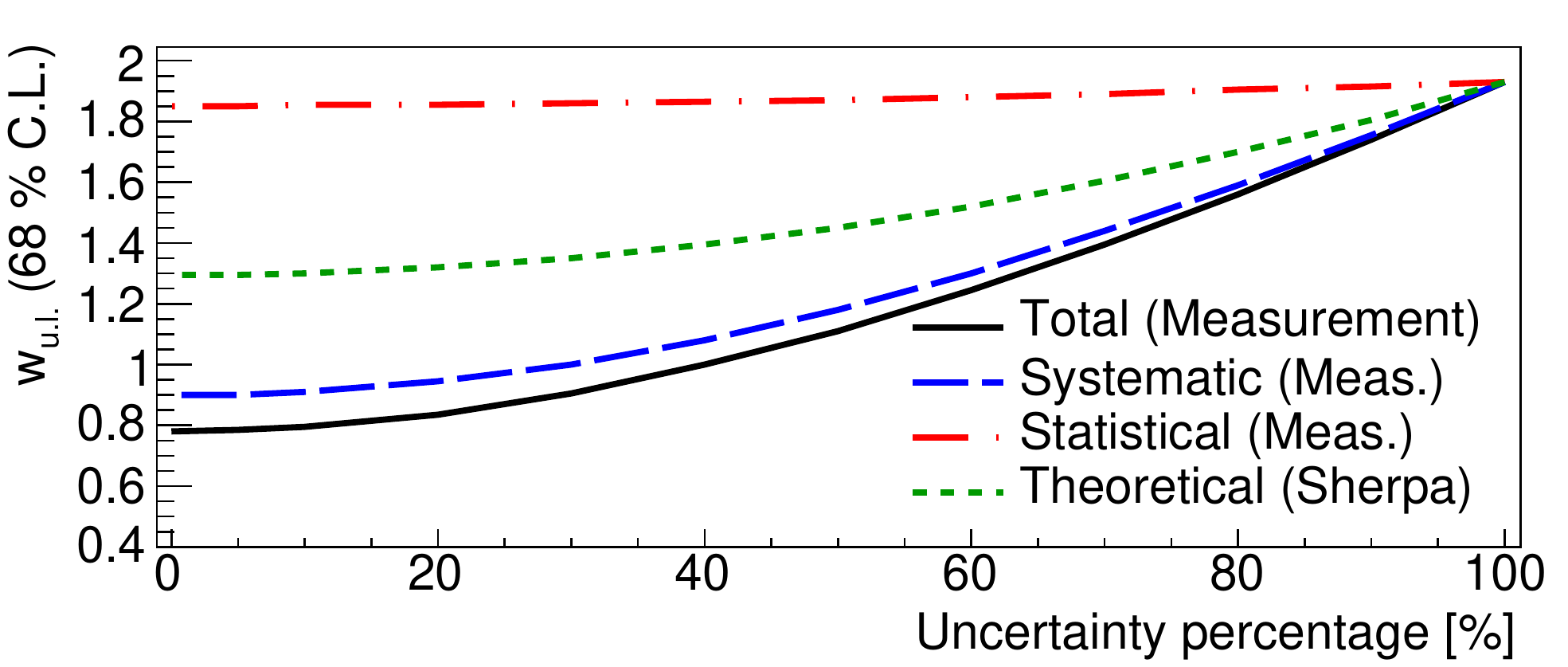}
  \caption{The dependence of the IC upper limit \wul{} at 68\% C.L. on the
           uncertainty percentage of the particular uncertainty
           component.}\label{fig4}
\end{figure}

The precision is limited by the experimental systematic
uncertainties --- mainly, by the $c$-tagging uncertainty which is predominantly
connected with the light jet scaling factors~\cite{Aaboud:2017skj}. It is also
limited by theoretical QCD scale uncertainties. The PDF uncertainties are
included in the predictions using the \sherpa{} (NLO). 
In contrast to these uncertainties, the statistical uncertainty
does not play a large role. All this can be seen in Fig.~\ref{fig4}, where the
dependence of the allowed IC upper limit \wul{} on different components of
uncertainty is shown. The allowed upper limit is presented four times, every
time the component of uncertainty in question is reduced from its actual value
(100~\%). This asssumes that the central values of the experiment does not
change. In order to obtain more reliable information on the IC probability in
the proton from future LHC data at $\sqrt{s} = 13$~\tev{} it is needed to have a
more realistic estimate of the theoretical scale uncertainties and reduce the
systematic uncertainties.

This problem can be, in fact, eliminated by employing the ``principle of maximum
conformality'' (PMC)~\cite{Brodsky:2013vpa} which sets renormalization scales by
shifting the $\beta$ terms in the pQCD series into the running coupling. The PMC
predictions are independent of the choice of renormalization scheme --- a key
requirement of the renormalization group. Its utilization will be the next step
of our study.

\noindent In summary:\\
A first estimate of the intrinsic charm probability in the proton has been
carried out utilizing recent ATLAS data on the prompt photon production
accompanied by the $c$-jet at $\sqrt{s} = 8$~\tev~\cite{Aaboud:2017skj}.  We
estimate the upper limit of the IC probability in proton about 1.93\%.  In order
to obtain more precise results on the intrinsic charm contribution one needs
additional data and at the same time reduced systematic uncertainties  which
come primarily from $c$-jet tagging. In particular, measurements of cross
sections of $\gamma + c$ and $\gamma + b$ production in $pp$-collisions at
$\sqrt{s} = 13$~\tev{} at high transverse momentum with high
statistics~\cite{Lipatov:2016feu} will be very useful since the ratio of photon
+ charm to photon + bottom cross sections is very sensitive to the IC
signal~\cite{Lipatov:2016feu, Brodsky:2016fyh}. The ratio, when \ety{} grows,
decreases in the absence of the IC contribution and stays flat or increases when
the IC contribution is included. Furthermore, measurements of $Z/W+c/b$
production in $pp$ collision at 13~\tev{} could also give additional significant
information on the intrinsic charm contribution~\cite{Beauchemin:2014rya,
Lipatov:2016feu,Brodsky:2016fyh,Lipatov:2018oxm}. Our study shows that the most
important source of theoretical uncertainty on \wcc, from the theory point of
view, is the dependence on the renormalization and factorization scales. This
can be reduced by the application of the Principle of Conformality (PMC), which
produces scheme-independent results, as well the calculation of the NNLO pQCD
contributions. Data at different energies at the LHC which checks scaling
predictions and future improvements in the accuracy of flavor tagging will be
important. These advances, together with a larger data sample (more than
100~fb$^{-1}$) at 13~\tev, should provide definitive information from the LHC on
the contribution of the non-perturbative intrinsic heavy quark contributions to
the fundamental structure of the proton. 

\noindent \textbf{Acknowledgements}\\
We thank A.A.~Glasov, R.~Keys, E.V.~Khramov, S.~Prince and S.M.~Turchikhin for
very helpful discussions. The authors are also grateful to S.P.~Baranov,
P.-H.~Beauchemin, I.R.~Boyko, Z.~Hubacek, H.~Jung, F.~Hautmann, B.~Nachman,
P.M.~Nadolsky, H.~Tores and L.~Rotali, N.A.~Rusakovich for useful discussions
and comments. The \sherpa{} calculations by J.S. were done at SIVVP, ITMS
26230120002 supported by the Research \& Development Operational Programme
funded by the ERDF\@. J.S. is thankful for the opportunity of using this
cluster. The work of A.V.L and M.A.M was supported in part by the grant of the
President of Russian Federation NS-7989.2016.2. A.V.L. and M.A.M. are also
grateful to DESY Directorate for the support in the framework of Moscow ---
DESY project on Monte-Carlo implementation for HERA --- LHC\@. M.A.M. was
supported by a grant of the Foundation for the Advancement of Theoretical
physics ``Basis{}'' 17--14--455--1 and RFBR grant 16--32--00176-mol-a. SJB is
supported by the U.S. Department of Energy, contract DE--AC02--76SF00515, the
SLAC Pub number is SLAC-PUB-17198 with title: Constraints on the Intrinsic Charm
Content of the Proton from Recent ATLAS Data.

\bibliography{bibfile}

\end{document}